\documentclass[twocolumn,prb,showpacs]{revtex4}

\usepackage{graphicx}
\usepackage{dcolumn}
\usepackage{amsmath2000}
\usepackage{amssymb}
\usepackage{xspace}
\usepackage{bm}

\newcommand{\Eqref}[1]{Eq.~\ref{#1}}
\newcommand{\Figref}[1]{Fig.~\ref{#1}}
\newcommand{\Citeref}[1]{Ref.~\onlinecite{#1}}

\renewcommand{\d}[1]{d \:\! #1}

\newcommand{\bvec}{\bm}
\newcommand{\maal}[1]{\:\text{#1}}
\renewcommand{\vec}{\bm}
\newcommand{\mat}{\mathsf}
\newcommand{\vk}{\vec{\kappa}}

\begin{document}

\title{Fermion-pairing on a square lattice in extreme magnetic fields}
\author{Sjur Mo}
 \email{Sjur.Mo@phys.ntnu.no}
\author{Asle Sudb{\o}}
 \email{Asle.Sudbo@phys.ntnu.no}       
\affiliation{%
  Department of Physics \\
  Norwegian University of Science and Technology, \\
  N-7491 Trondheim, Norway \\} 
\date{\today}

\begin{abstract}
  
  We consider the Cooper-problem on a two-dimensional, square lattice
  with a uniform, perpendicular magnetic field. Only rational flux
  fractions are considered. An extended (real-space) Hubbard model
  including nearest and next nearest neighbor interactions is
  transformed to ``$k$-space", or more precisely, to the space of
  eigenfunctions of Harper's equation, which constitute basis
  functions of the magnetic translation group for the lattice.  A
  BCS-like truncation of the interaction term is performed. Expanding
  the interactions in the basis functions of the irreducible
  representations of the point group $C_{4\nu}$ of the square lattice
  simplify calculations.  The numerical results indicate enhanced
  binding compared to zero magnetic field, and thus re-entrant
  superconducting pairing at extreme magnetic fields, well beyond the
  point where the usual semi-classical treatment of the magnetic field
  breaks down.

\end{abstract}

\maketitle

\section{Introduction}

The relationship between an external magnetic field and
superconductivity is very important both from a theoretical and a
practical point of view.  It is well known that superconductivity will
disappear above a certain magnetic field due to orbital frustration
within a semi-classical treatment of the problem, i.e. the pairing
states $(-\vec{k},\vec{k})$ are not good quantum numbers.  On the
other hand, it was realized some years ago that there may be
additional interesting effects as a magnetic field is increased to
large values at low temperatures.  Rasolt and
Te{\v{s}}anovi{\'{c}}~\cite{Rasolt:1992} showed that re-entrant
superconductivity may appear for fields above $H_{c2}$ due to Landau
level quantization. Because electrons in Landau levels will not be
influenced by orbital frustration in a magnetic field, an increased
magnetic field may in fact enhance superconductivity. The re-entrance
has no direct relationship with the low field superconductivity, and
some materials may even be superconducting \emph{only} in high
magnetic fields! When Zeeman splitting was incorporated the re-entrant
superconducting phase disappeared at high enough fields.  We will
investigate if a similar phenomenon also will happen when a periodic
lattice is incorporated, modifying the Landau level picture of the
single-particle spectrum to Hofstadter bands.~\cite{Hofstadter:1976}
The influence of Zeeman splitting (i.e.  effective Land{\'e} $g$-factor)
on the Cooper binding energy, will also be considered.

In order to do this, we have considered the Cooper-problem on a
two-dimensional, square lattice with a uniform perpendicular magnetic
field, i.e. we consider pairing of fermions in large magnetic fields
at zero temperature. The same problem in zero field has been studied
in \Citeref{Otnes:1997}. The model has on-site, nearest neighbor, and
next-nearest neighbor interactions, where the values used for the
interactions are \emph{input} to the model. We do not have in mind any
particular microscopic mechanism causing these effective interactions,
but assume that they can be attractive.  Since the interaction
potential has finite range, it is possible to expand the Fourier
transformed version in a finite number of functions. These are basis
functions of the irreducible representations of the point group
$C_{4\nu}$ of the square lattice.

In the absence of interactions, the Schr{\"o}dinger equation for the
problem is
the well known Harper's equation, whose spectrum and eigenfunctions
have been extensively studied, see e.g.~\Citeref{Harper:1955,Brown:1964,Brown:1968b,%
Zak:1964a,Zak:1964b,Zak:1964d,Zak:1965,Dana:1983,Dana:1985,Hofstadter:1976,%
Obermair:1981a,Obermair:1981b,Thouless:1982,Schellnhuber:1982,Hasegawa:1989}.
The spectrum is the so called Hofstadter's butterfly. If the ratio of
the flux per plaquette to the flux quantum is a rational number
($\frac{\phi}{\phi_0} = \frac{p}{q}$), the energy bands split into $q$
sub-bands. For irrational ratios the spectrum is a Cantor
set.~\cite{Hofstadter:1976} Effects of electron correlations on the
Hofstadter spectrum have been studied by several authors, see
e.g.~\Citeref{Barelli:1996,Shepelyansky:1996,Anh:1997,Doh:1998,Hong:1999,Hong:2000},
however they do not consider the possibility of Cooper pairing through an 
attractive effective interaction.

For low magnetic fields the behavior of the system can be estimated by
treating the field in a perturbative manner. Using the zero field
correlation length one can find a value for $H_{c2}$.  The limitations
of this approach is that it neglects the change in the wave-functions
of the Cooper pairs due to the magnetic field. For very large magnetic
fields, a more appropriate approach to the problem is to solve the
single-particle spectrum first in the presence of the magnetic field,
and then introduce the pairing interaction as a singular perturbation
on this problem. If one is primarily interested in the behavior near
$T_c(H)$, one can exploit the fact that the order parameter/energy gap
is small and develop linearized formulas as done in
e.g.~\Citeref{Mierzejewski:1999prb,Maska:2000,Miyazaki:1999cm}.

We transform the extended Hubbard model (real-space) to the space of
eigenfunctions of the Harper's equation which diagonalize the single
electron part. A BCS-like truncation is then performed and the
interactions are expanded in the basis functions of the point group
$C_{4\nu}$ of the square lattice. {}From this Hamiltonian, an exact
solution of the Cooper problem is in principle possible to obtain,
however one encounters the problem that the associated matrix problem
increases dramatically upon a reduction of the magnetic field.  In
fact, the matrix dimensions will be $\sim q^2 \times q^2$ and only
fields far above experimentally achievable ones can be handled in
explicit numerical solutions at present.\footnote{E.g. a lattice
constant $a = 4\maal{\AA} $ and $B=100\maal{T}$ gives $q \sim 250$
or $q^2 \sim 10^5$.} We will consider extremely high fields where
the flux per plaquette is a sizeable fraction of $\phi_0$, typically
$q<8$.

\section{Theory}

\subsection{The model}

We use an extended Hubbard model with nearest and next-nearest
neighbor hopping matrix elements $t$ and $t'$ respectively, given by
\begin{subequations}
\begin{equation}
        H = H_0 + H_I,
\end{equation}
where
\begin{align} 
        H_0 &= \sum_{i,\sigma} [\epsilon(\sigma)-\mu]
c^{\dagger}_{i,\sigma}  c_{i,\sigma}
                 \nonumber \\ 
            &- t\sum_{\langle i,j \rangle,\sigma} e^{\phi_{ij}} 
                                c^{\dagger}_{i,\sigma}  c_{j,\sigma} 
             - t'\sum_{\langle\langle i,j \rangle\rangle,\sigma}
e^{\phi_{ij}} 
                        c^{\dagger}_{i,\sigma}  c_{j,\sigma} \label{H0} \\
\intertext{and}
        H_I &= \frac{U}{2} \sum_{i,\sigma}  c^{\dagger}_{i,\sigma} 
c_{i,\sigma} 
                               c^{\dagger}_{i,-\sigma}  c_{i,-\sigma}
                \nonumber \\
            &+ \frac{V}{2} \sum_{\langle i,j \rangle,\sigma,\sigma'} 
                               c^{\dagger}_{i,\sigma}  c_{i,\sigma} 
                               c^{\dagger}_{j,\sigma'} c_{j,\sigma'}
                \nonumber \\ 
            &+ \frac{W}{2} \sum_{\langle\langle i,j
\rangle\rangle,\sigma,\sigma'} 
                               c^{\dagger}_{i,\sigma}  c_{i,\sigma} 
                               c^{\dagger}_{j,\sigma'} c_{j,\sigma'}
\label{HI}.
\end{align}
\end{subequations}
Here $U$, $V$ and $W$ are effective interactions between electrons at
the same site, nearest neighbor sites and next nearest neighbor sites,
respectively. The noninteracting model including next nearest neighbor
hopping has been studied in e.g.~\Citeref{Han:1994}. The Zeeman
splitting is given by $\epsilon(\sigma) = - g_s \mu_B B \sigma/2$
where $\sigma = +(-) [=\uparrow(\downarrow)]$,
$g_s$ is Land{\'e} $g$-factor, and $\mu$ is the chemical potential.%
\footnote{%
  In an external magnetic field $|\epsilon(-\sigma)-\epsilon(\sigma)|
  = g_s \frac{e \hbar}{2 m} B = 2\pi g^{\ast} \frac{p}{q} t$ where
  $m^{\ast}$ is the effective mass and $g_s=2$ in most circumstances.
  This is found by using that $\frac{\hbar^2}{2 m^{\ast} a^2} \equiv
  t$, $\frac{2\pi B a^2}{h/e} = 2 \pi \frac{\phi}{\phi_0}$ and $g_s
  \frac{m^{\ast}}{m} \equiv g^{\ast}$.  } %
The Peierls phase factors~\cite{Peierls:1933} for hopping from lattice
site $j$ to $i$ is
\begin{equation}
        \phi_{ij} = -\frac{2 \pi}{\phi_0} \int^i_j \bvec{A} \cdot
\d{\bvec{\ell}}
\end{equation}
where $\phi_0 = h/e$.  Employing a Landau-gauge
$\bvec{A}=Bx\hat{e}_y$, introducing $g=2\pi p/q = 2\pi Ba^2 /\phi_0$,
and setting $a=1$, the phase factor can be written as
\begin{equation} 
\label{phase}
        \phi_{ij} = \begin{cases}
                        0 & \text{if}\quad |\bvec{r}_i-\bvec{r}_j| = m\hat{e}_x \\
                        g \frac{x_i + x_j}{2} n & \text{if}\quad 
                        |\bvec{r}_i-\bvec{r}_j| = m\hat{e}_x +n\hat{e}_y. 
                    \end{cases}
\end{equation}
Using this, and writing the wave function as (we have chosen the
normalization $\sum_{i=1}^N |u_{\mu,\nu,\ell}(x_i)|^2 = 1$ where $N$
is the number of lattice sites.)
\begin{equation}
        \psi_{\mu,\nu,\ell}(x_m,y_m) = \langle \bvec{r}_m | \vk , l \rangle
        = e^{i(\mu x_m + \nu y_m)} u_{\mu,\nu,\ell}(x_m),
\end{equation}  
where we have introduced $\vk = (\mu,\nu)$, the Schr{\"{o}}dinger
equation can be written as
\begin{gather}
\label{Harper}
        -e^{ i\mu}\left\{t+2t'\cos[g(m+1/2)+\nu)] \right\}
u_{\mu,\nu,\ell}(x_{m+1}) \notag \\
        -e^{-i\mu}\left\{t+2t'\cos[g(m-1/2)+\nu)] \right\}
u_{\mu,\nu,\ell}(x_{m-1}) \notag \\
        -\left\{ 2 t \cos[gm+\nu] \right\} u_{\mu,\nu,\ell}(x_m)   
        = \epsilon_{\mu,\nu,\ell} u_{\mu,\nu,\ell}(x_m).
\end{gather}
Here, $x$ is taken modulo $q$ due to the periodicity for $m\to m+q$, such
that
 the functions $u$ constitute the periodic part of Bloch functions
on the \emph{magnetic} lattice. The equation can be written as a
$q\times q$ matrix. It is furthermore easily seen that the equation is
periodic
for $\nu\to\nu+2\pi$. By introducing
$\hat{u}_{\mu,\nu,\ell}(x_m)=e^{i\mu x_{m\bmod q}}
u_{\mu,\nu,\ell}(x_m)$ it is easily shown that the equation for
$\hat{u}_{\mu,\nu,\ell}(x_m)$ is periodic for $\mu\to\mu+2\pi/q$.
As a consequence of the fact that  $x_m - x_{m \bmod q} = n q$, it is clear
that
$\psi_{\mu,\nu,\ell}(x_m,y_m)$ must have the same periodicity.  The
different eigenvalues are numbered with $\ell$ and each value
corresponds to a different Harper- or Hofstadter-band.  Using the
completeness relation $1 =
\sum_{\vk,\ell} |\vk,\ell \rangle \langle \vk,\ell |$,
the creation/annihilation operators are seen to transform as%
\footnote{%
  We can introduce ``Wannier functions''or localized states $|i
  \rangle = \sum_{\vk,\ell} \psi^{\ast}_{\vk,\ell}(\bvec{r}_i)
  |\vk,\ell \rangle$.  }
\begin{equation}
\begin{split}
\label{transform}
        c^{\dagger}_{i,\sigma} &= \sum_{\ell=1}^{q} \sum_{\vk}
                          e^{-i \vk \cdot \vec{r}_i} 
                          u^{\ast}_{\vk,\ell}(x_i)
                          c^{\dagger}_{\vk,\ell,\sigma} \\
        c_{i,\sigma} &= \sum_{\ell=1}^{q} \sum_{\vk}
                          e^{i \vk \cdot \vec{r}_i} 
                          u_{\vk,\ell}(x_i)
                          c_{\vk,\ell,\sigma}
\end{split}
\end{equation}
where $x_i = x_i \bmod q$.
Inserting \Eqref{transform} in \Eqref{H0} we get
\begin{equation}
        H_0     = \sum_{\vk,\ell,\sigma}
                \left[ \epsilon_{\vk,\ell} + \epsilon(\sigma)-\mu \right]
                c^{\dagger}_{\vk,\ell,\sigma} c_{\vk,\ell,\sigma}.
\end{equation}
The two-particle terms in \Eqref{HI} are all  of the form
\begin{equation}
        \frac{1}{2}
        \sum_{i,j,\sigma,\sigma'} V(|i-j|) 
         c^{\dagger}_{i,\sigma}  c_{i,\sigma}
         c^{\dagger}_{j,\sigma'} c_{j,\sigma'}. 
\end{equation}
By using \Eqref{transform} and a form of BCS-truncation of the interaction
term
(See appendix~\ref{sec-HIdet} for details)
\begin{equation}
        \begin{array} {cccc}
                \sigma' &=& -\sigma & \\
                \vk' &=& -\vk, & (-\pi/q,-\pi)\leq \vk < (\pi/q,\pi),
        \end{array}
\end{equation}
the two-particle terms can be transformed into
\begin{equation}
        \frac{1}{2}
        \sum_{\ell_1,\ell_2,\ell'_1,\ell'_2} 
        \sum_{\vk,\vk', \sigma}
        V^{\ell_1 \ell_2,\ell'_1 \ell'_2}_{\vk,\vk'}
                 c^{\dagger}_{\vk,\ell_1,\sigma}
c^{\dagger}_{-\vk,\ell_2,-\sigma} 
                 c_{\vk',\ell'_1,\sigma} c_{-\vk',\ell'_2,-\sigma}, 
\end{equation}
where
\begin{equation} \label{Vkk}
        V^{\ell_1 \ell_2,\ell'_1 \ell'_2}_{\vk,\vk'}
        = \sum_{\vec{G},\vec{G}'} 
          \sum_{\eta=1}^{5}
          \lambda_{\eta} 
          \widetilde{B}^{\ast}_{\eta,\vec{G},\vec{G}'}(\ell_1,\ell_2,\vk)
          \widetilde{B}_{\eta,\vec{G},\vec{G}'}(\ell'_1,\ell'_2,\vk').
\end{equation}
Here, we have introduced $\bvec{\lambda}=(U,V,W,V,W)$ and functions
$\widetilde{B}$ given by
\begin{subequations}
\begin{equation}
        \widetilde{B}_{\eta,\vec{G},\vec{G}'}(\ell_1,\ell_2,\vk)
        =
        \widetilde{u}_{\vk,\ell_1}(\vec{G})
                        \widetilde{u}_{-\vk,\ell_2}(\vec{G}')B_{\eta}(\vk),
\end{equation}
where
\begin{align}
        B_1(\vk) &= \sqrt{N}                                            
\nonumber \\
        B_2(\vk) &= \sqrt{N} \left[\cos(\kappa_x)+\cos(\kappa_y)\right] 
\nonumber \\
        B_3(\vk) &= \sqrt{2 N} \cos(\kappa_x) \cos(\kappa_y)               
       \\
        B_4(\vk) &= \sqrt{N} \left[\cos(\kappa_x)-\cos(\kappa_y)\right] 
\nonumber \\
        B_5(\vk) &= \sqrt{2 N} \sin(\kappa_x)\sin(\kappa_y),            
\nonumber \\
\intertext{and}
        \widetilde{u}_{\vk,\ell}(\vec{G}) &= \frac{1}{\sqrt{q}}
        \sum_{i=1}^q u_{\vk,\ell}(x_i) e^{-i \vec{G} x_i}.
        \label{u(G)}
\end{align}
\end{subequations}
Here $\vec{G}=(G_x,G_y)$, where $G_x = 0, 2\pi/q, \cdots (q-1)2\pi/q$
is a reciprocal lattice vector for the magnetic lattice, and $G_y=0$.

\subsection{The Cooper-problem}

The two-particle Schr{\"{o}}dinger equation is given by
\begin{equation} \label{Sch1}
        (H_0 + H_I)|1,2 \rangle
        = E |1,2 \rangle. 
\end{equation}
Without interactions the Schr{\"{o}}dinger equation is
\begin{equation}
\begin{split}
        H_0 |\vk,&\ell_1,\sigma ; -\vk,\ell_2,-\sigma \rangle_0 \\
        &= ( \epsilon_{\vec{ \kappa},\ell_1,\sigma} 
        +\epsilon_{-\vk,\ell_2,-\sigma})
        |\vk,\ell_1,\sigma ; -\vk,\ell_2,-\sigma \rangle_0,
\end{split}
\end{equation}
and it is possible to expand the two-particle states 
in the complete set $|\; \rangle_0$:
\begin{eqnarray}
        |1,2 \rangle
        = \sum_{\vk,\ell_1,\ell_2,\sigma} 
        \theta(\epsilon_{\vec{ \kappa},\ell_1,\sigma}-\epsilon_F)
        \theta(\epsilon_{-\vec{ \kappa},\ell_2,-\sigma}-\epsilon_F)
        \nonumber \\
        \times
        a^{\ell_1,\ell_2}_{\kappa,\sigma} 
        |\vk,\ell_1,\sigma ; -\vk,\ell_2,-\sigma \rangle_0,
\end{eqnarray}
where $\epsilon_F$ is the Fermi-level and $\theta(x<0)=0,\;
\theta(x>0)=1$.  Inserting this in the Schr{\"{o}}dinger equation
\Eqref{Sch1}, multiplying from the left by ${}_0 \langle
\vk',\ell'_1;-\vk',\ell'_2|$, using \Eqref{Vkk}, and introducing the
shorthand notations $ \chi = \left\{ \vk,\ell_1,\ell_2 \right\}$ and
$\xi = \left\{ \eta,\vec{G},\vec{G}' \right\}$ give
\begin{equation}
\begin{split}
        \sum_{\xi} 
                \widetilde{B}^{\ast}_{\xi}(&\chi') 
                \underbrace{\sum_{\chi>\chi_F} \lambda_{\xi}
\widetilde{B}_{\xi}(\chi)
                a_{\chi,\sigma'}}_{\equiv A_{\xi,\sigma'}}
        \\
        & =
        a_{\chi',\sigma'}
        \left[ E - (\epsilon_{\vec{ \kappa'},\ell'_1,\sigma'} 
                 +  \epsilon_{-\vec{\kappa'},\ell'_2,-\sigma'})
        \right], 
\end{split}
\end{equation}
which implies
\begin{equation}
        a_{\chi',\sigma'} 
        = \frac{ \sum_{\xi}\widetilde{B}^{\ast}_{\xi}(\chi')
A_{\xi,\sigma'}}
               {E - ( \epsilon_{ \vec{\kappa'},\ell'_1, \sigma'} 
                     +\epsilon_{-\vec{\kappa'},\ell'_2,-\sigma'})}
        ,\quad \chi' > \chi_F.
\end{equation}
Here the notation $\chi' > \chi_F$ means $\epsilon_{\vec{
    \kappa'},\ell'_1,\sigma'} > \epsilon_F \quad \text{and} \quad
\epsilon_{-\vec{ \kappa'},\ell'_2,-\sigma'} > \epsilon_F$.  Inserting
the expression for $a_{\chi',\sigma'}$ into \Eqref{Sch1}, multiplying
by $\widetilde{B}_{\xi'}(\chi')$ and summing over $\chi' > \chi_F$
gives
\begin{eqnarray} 
        \sum_{\xi''} 
        \underbrace{
                \sum_{\xi} 
                        \lambda_{\xi} 
                        \Gamma_{\xi',\xi}
                        M_{\xi'',\xi}(E,\sigma')        
        }_{\equiv D_{\xi',\xi''}(E,\sigma')}
        A_{\xi'',\sigma'}
        =
        \sum_{\xi} \Gamma_{\xi',\xi} A_{\xi,\sigma'},
\end{eqnarray}
where
\begin{equation}
        \Gamma_{\xi',\xi}
        \equiv 
        \sum_{\chi'>\chi_F}
                        \widetilde{B}^{\ast}_{\xi}(\chi')
                        \widetilde{B}_{\xi'}(\chi'),
\end{equation}
and
\begin{equation}
        M_{\xi'',\xi}(E,\sigma')
        =
        \sum_{\chi>\chi_F} 
                \frac{\widetilde{B}^{\ast}_{\xi''}(\chi)\widetilde{B}_{\xi}(\chi)}
                        {E-( \epsilon_{ \vk,\ell_1, \sigma'}
                            +\epsilon_{-\vk,\ell_2,-\sigma'})}.
\end{equation}
The final equation is
\begin{equation} 
        \sum_{\xi} D_{\xi',\xi}(E,\sigma')A_{\xi,\sigma'}
        = \sum_{\xi} \Gamma_{\xi',\xi} A_{\xi,\sigma'},
\end{equation}
which has a (nontrivial) solution if and only if
$\det(\mat{D}(E)-\mat{\Gamma})=0$.  This determines $E$ and we can
find the vector components $A_{\xi,\sigma'}$, and thereby the Cooper
wave functions in $(\vk,\ell)$-space,
$a^{\ell_1,\ell_2}_{\vk,\sigma'}$. The real space wave function is
then
\begin{eqnarray}
        \langle ij | 1,2 \rangle
        = \sum_{\vk,\ell_1,\ell_2,\sigma} 
        \theta(\epsilon_{\vec{ \kappa},\ell_1,\sigma}-\epsilon_F)
        \theta(\epsilon_{-\vec{ \kappa},\ell_2,-\sigma}-\epsilon_F)
        \nonumber \\
        \times
        a^{\ell_1,\ell_2}_{\kappa,\sigma} 
        \underbrace{
                \langle i |\vk,\ell_1,\sigma   \rangle_0
                \langle j |-\vk,\ell_2,-\sigma \rangle_0
        }_{ = u_{\vk,\ell_1}(x_i) u_{-\vk,\ell_2}(x_j)
                 e^{i \vk \cdot (\vec{r}_i - \vec{r}_j)}},
\end{eqnarray} 
or in mass-center and relative coordinates
$X=(x_i+x_j)/2,\quad\vec{\rho}= \vec{r}_i-\vec{r}_j$:
\begin{multline}
        \langle X,\vec{\rho} | 1,2 \rangle
        = \sum_{\vk,\ell_1,\ell_2,\sigma} 
        \theta(\epsilon_{\vec{ \kappa},\ell_1,\sigma}-\epsilon_F)
        \theta(\epsilon_{-\vec{ \kappa},\ell_2,-\sigma}-\epsilon_F)
        \\
        \times
        a^{\ell_1,\ell_2}_{\kappa,\sigma} 
        u_{\vk,\ell_1}(X+\frac{\rho_x}{2})
u_{-\vk,\ell_2}(X-\frac{\rho_x}{2})
                 e^{i \vk \cdot \vec{\rho}}.
\end{multline}

\section{Results and discussion}

In the previous section we have developed a framework for computations
involving both nearest- and next nearest neighbor hopping and on-site,
nearest neighbor and next nearest neighbor electron interactions.  As
a first step, we have studied explicit numerical solutions for the
situation with only on-site \emph{attractive} potential. This
corresponds to setting $U<0$ and $V=W=0$ in the above formulas, i.e.
the Hubbard model which in the absence of a magnetic field is
expected to exhibit s-wave superconductivity.  The next nearest
neighbor hopping term $t'$ is also set to zero.

\begin{figure}[htbp]
  \centering 
  \includegraphics[angle=270,width=3.4in]{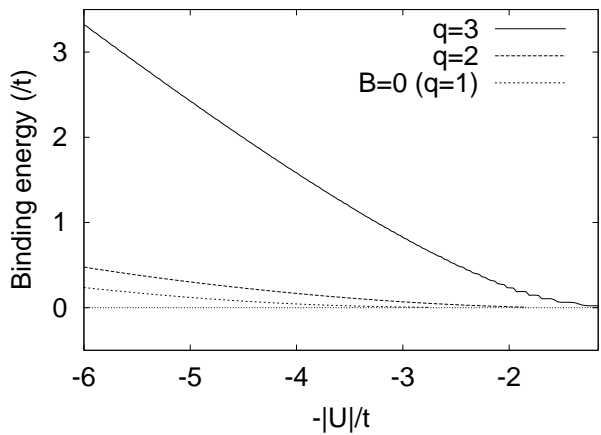}
  \caption{ Binding energy for Cooper pair as function of on-site 
    effective potential for different magnetic fields ($q$).  The
    Zeeman spin splitting is neglected and the band filling is $n=0.4$
    ($0.5$ is half filled band).
    \label{fig1}
  }  
\end{figure}

We have calculated the Cooper-pair binding energy as function of
$-|U|$ for different values of the band filling $n$. A typical result
for $n=0.4$ is shown in \Figref{fig1}.  The results show that the
binding energy is an increasing function of $|U|$ both for $q=1$
(corresponds to zero field) and for $q=2, 3$. Neglecting
spin-splitting, the binding energy is clearly higher for $q>1$ and is
an increasing function of $q$ for the extreme fields we are
considering. An increase in $q$ makes the allowed energy bands
narrower, and thereby increases the density of states, and hence this
result is to be expected. Since the Fermi level normally lies in a
band, the density of states at the Fermi level is higher than in zero
field.

\begin{figure}[htbp]
  \centering 
  \includegraphics[angle=270,width=3.4in]{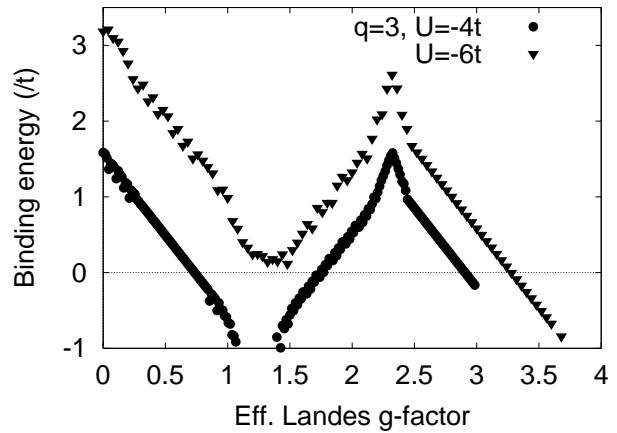}
  \caption{ Binding energy for Cooper pair as function of the effective
    Land{\'e} $g$-factor (spin-splitting). The band filling is $n=0.4$
    and $q=3$.
    \label{fig2}
  }
\end{figure}

When one includes the Zeeman spin splitting, the Hofstadter bands will
split. As the effective Land{\'e} $g$-factor increases, the spin up and
spin down bands will move relative to each other, and ultimately be
completely separated.  A typical result for $U=-4t$ is shown in
\Figref{fig2}.  The binding energy is not a monotonically
decreasing function, but increases and decreases as the bands overlap
more and less. When the bands have passed each other and the spin up
states are increasingly separated from the spin down states the
binding energy is a decreasing function of the $g$-factor.

\begin{figure}[htbp]
  \centering 
  \includegraphics[angle=270,width=3.4in]{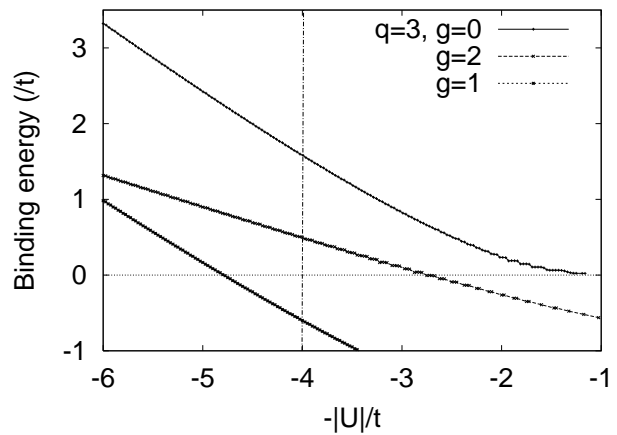}
  \caption{ Binding energy for Cooper pair as function of on-site 
    effective potential for different values of the effective
    Land{\'e} $g$-factor for band filling $n=0.4$ and $q=3$.
    \label{fig3}
  }
\end{figure}

A similar plot as in \Figref{fig1} is shown in \Figref{fig3}. Here the
binding energy as function of $-|U|$ is plotted for different values
of the spin splitting for $q=3$.

\section{Conclusion}

For the attractive Hubbard model in extreme magnetic fields the above
results indicate that the binding energy is an increasing function of
$q$, i.e the binding energy is higher in a magnetic field (extremely
strong) than in the zero field situation.  It is not clear what will
happen for even higher values of $q$, but there has to be a maximum
before the decrease towards the normal $H_{c2}$ from above.  If the
effective Land{\'e} $g$-factor is not too high, this result is still
valid when we include the Zeeman effect. The results tell us that in
the model we have considered, re-entrant superconductivity as a
function of magnetic field will appear if the on-site attractive
potential is sufficiently strong. For real materials, phase
transitions to different phases may appear for lower fields than those
we have studied.


\begin{acknowledgments}
  We thank Prof.~Z.~Te{\v s}anovi\'{c} for useful discussions. This
  work was supported by the Norwegian Research Council via the High
  Performance Computing Program and Grant No.~124106/410.
\end{acknowledgments}


\begin{widetext}
\appendix
\section{Transformation of the interaction terms} 
\label{sec-HIdet}

By using \Eqref{transform} and introducing $\vec{r}_j = \vec{r}_i +
\vec{\delta} $, we can write
\begin{equation}
\begin{split}
H_I &=  \frac{1}{2}
        \sum_{i,j,\sigma,\sigma'} V(|i-j|) 
         c^{\dagger}_{i,\sigma}  c_{i,\sigma}
         c^{\dagger}_{j,\sigma'} c_{j,\sigma'} \\
        &=
        \frac{1}{2}
        \sum_{\vk_1 \cdots \vk_4}
        \sum_{i,\vec{\delta}} V(|\vec{\delta}|) 
        e^{ i(\vk_4-\vk_3) \cdot \vec{\delta}}
        e^{ i[ (\vk_2+\vk_4)
              -(\vk_1+\vk_3)] \cdot \vec{r}_i} 
        \\
        & \quad \times \sum_{\ell_1 \cdots l_4}
                u^{\ast}_{\vk_1,\ell_1}(x_i)
                u_{\vk_2,\ell_2}(x_i)
                u^{\ast}_{\vk_3,\ell_3}(x_i + \delta_x)
                u_{\vk_4,\ell_4}(x_i + \delta_x) 
        \\
        & \quad \times \sum_{\sigma,\sigma'}
                c^{\dagger}_{\vk_1,\ell_1,\sigma}
                c_{\vk_2,\ell_2,\sigma}
                c^{\dagger}_{\vk_3,\ell_3,\sigma'}
                c_{\vk_4,\ell_4,\sigma'}.
\end{split}
\end{equation}
Since $u_{\vk,\ell}(x)$ is periodic with period $q$ in the
$x$-direction (and $1$ in the $y$-direction), we can use a
Fourier-transform over the \emph{magnetic Brillouin zone}
\begin{align}
        u_{\vk,\ell}(x_i) = \frac{1}{\sqrt{q}}
                \sum_{G_x} \widetilde{u}_{\vk,\ell}(G_x)
                e^{iG_x x_i},
\intertext{or}
        u_{\vk,\ell}(\vec{r}_i) = \frac{1}{\sqrt{q}}
                \sum_{\vec{G}} \widetilde{u}_{\vk,\ell}(\vec{G})
                e^{i\vec{G}\cdot\vec{r}_i},
\end{align}
where $G_x = 0, 2 \pi/q, \cdots (q-1)2\pi/q$ and $G_y = 0$. By using
this, we get
\begin{equation}
\begin{split}
H_I &=  \frac{1}{2 q^2}  
        \sum_{\substack{\vk_1 \cdots \vk_4 \\ \ell_1 \cdots \ell_4}}
        \sum_{\vec{G}_1 \cdots \vec{G}_4} 
                \widetilde{u}^{\ast}_{\vk_1,\ell_1}(\vec{G}_1)
                \widetilde{u}_{\vk_2,\ell_2}(\vec{G}_2)
                \times
                \widetilde{u}^{\ast}_{\vk_3,\ell_3}(\vec{G}_3)
                \widetilde{u}_{\vk_4,\ell_4}(\vec{G}_4)
        \sum_{\vec{\delta}}
                V(|\vec{\delta}|)
                e^{ i(\vk_4-\vk_3) 
                        \cdot \vec{\delta}}
                e^{ i(\vec{G}_4-\vec{G}_3) 
                        \cdot \vec{\delta}}
        \\
        &\quad \times
                \underbrace{
                        \sum_{i}
                        e^{ i[ (\vk_2+\vk_4)-(\vk_1+\vk_3)]\cdot \vec{r}_i}
                        e^{ i[ (\vec{G}_2+\vec{G}_4)-(\vec{G}_1+\vec{G}_3)]\cdot\vec{r}_i}
                }_{\equiv N\delta_{(\vk_2+\vec{G}_2)+(\vk_4+\vec{G}_4),  
                (\vk_1+\vec{G}_1)+(\vk_3+\vec{G}_3)}}
                \sum_{\sigma,\sigma'} 
                c^{\dagger}_{\vk_1,\ell_1,\sigma}
                c_{\vk_2,\ell_2,\sigma}
                c^{\dagger}_{\vk_3,\ell_3,\sigma'}
                c_{\vk_4,\ell_4,\sigma'}.
\end{split}
\end{equation}
If we then define $\vk_q$ by
\begin{equation}
        (\vk_1+\vec{G}_1)-(\vk_2+\vec{G}_2)
        =
        (\vk_4+\vec{G}_4)-(\vk_3+\vec{G}_3)
        \equiv \vk_q,           
\end{equation}
and 
\begin{equation}
        \widetilde{V}(\vk_q) \equiv N \sum_{\vec{\delta}} V(|\vec{\delta}|)
                                         e^{i \vk_q \cdot \vec{\delta}},   
\end{equation}
we can write
\begin{multline}
H_I =   \frac{1}{2 q^2}
        \sum_{\substack{\vk_2,\vk_3,\vk_q \\ \ell_1 \cdots \ell_4}}
        \widetilde{V}(\vk_q) 
        \sum_{\vec{G}_1 \cdots \vec{G}_4}
                \widetilde{u}^{\ast}_{\vk_2+\vk_q
                               -(\vec{G}_1-\vec{G}_2),\ell_1}(\vec{G}_1)
                \widetilde{u}_{\vk_2,\ell_2}(\vec{G}_2)
                \widetilde{u}^{\ast}_{\vk_3,\ell_3}(\vec{G}_3)
                \widetilde{u}_{\vk_3+\vk_q
                           -(\vec{G}_4-\vec{G}_3),\ell_4}(\vec{G}_4)
        \\
                \times
                \sum_{\sigma,\sigma'} 
                c^{\dagger}_{\vk_2+\vk_q
                             -(\vec{G}_1-\vec{G}_2) ,\ell_1,\sigma}
                c_{\vk_2,\ell_2,\sigma}
                c^{\dagger}_{\vk_3,\ell_3,\sigma'}
                c_{\vk_3+\vk_q
                   -(\vec{G}_4-\vec{G}_3),\ell_4,\sigma'}.
\end{multline}
This can be written on a more symmetric form by letting
$\vk_2\to\vk - \frac{\vk_q}{2}, \vk_3\to\vk' - \frac{\vk_q}{2}$:
\begin{multline}
H_I =   \frac{1}{2 q^2} 
        \sum_{\substack{\vk,\vk',\vk_q \\ \ell_1 \cdots \ell_4}}
        \widetilde{V}(\vk_q) 
        \sum_{\vec{G}_1 \cdots \vec{G}_4}
                \widetilde{u}^{\ast}_{\vk+\frac{\vk_q}{2}
                               -(\vec{G}_1-\vec{G}_2),\ell_1}(\vec{G}_1)
                \widetilde{u}_{\vk-\frac{\vk_q}{2},\ell_2}
                          (\vec{G}_2)
                \widetilde{u}^{\ast}_{\vk'-\frac{\vk_q}{2},\ell_3}
                          (\vec{G}_3)
                \widetilde{u}_{\vk'+\frac{\vk_q}{2}
                           -(\vec{G}_4-\vec{G}_3),\ell_4}(\vec{G}_4)
        \\
                \times
                \sum_{\sigma,\sigma'} 
                c^{\dagger}_{\vk+\frac{\vk_q}{2}
                             -(\vec{G}_1-\vec{G}_2) ,\ell_1,\sigma}
                c_{\vk-\frac{\vk_q}{2},\ell_2,\sigma}
                c^{\dagger}_{\vk'-\frac{\vk_q}{2},\ell_3,\sigma'}
                c_{\vk'+\frac{\vk_q}{2}
                   -(\vec{G}_4-\vec{G}_3),\ell_4,\sigma'}.
\end{multline}
Now we perform a BCS like truncation
\begin{equation}
\begin{split}
        \sigma' &= -\sigma \\
        \vk'    &= -\vk    ;\qquad (-\pi/q,-\pi) \leq \vk < (\pi/q,\pi),
\end{split}
\end{equation}
where $\vk$ is in the magnetic Brillouin zone, i.e. we assume that the
Cooper pair is a spin singlet. Due to the antisymmetry of the spin
part of the wave-function, the space part has to be symmetric.  If
$\vk'+\vk \neq 0$, the wave-function for the pairs center of mass
would have a modulation. This modulation will generally be
incommensurable with the underlying lattice. Thus
\begin{multline}
H_I =   \frac{1}{2 q^2} 
        \sum_{\substack{\vk,\vk_q,\sigma \\ \ell_1 \cdots \ell_4}}
        \widetilde{V}(\vk_q) 
        \sum_{\vec{G}_1 \cdots \vec{G}_4} 
                \widetilde{u}^{\ast}_{\vk+\frac{\vk_q}{2}
                               -(\vec{G}_1-\vec{G}_2),\ell_1}(\vec{G}_1)
                \widetilde{u}_{\vk-\frac{\vk_q}{2},\ell_2}
                          (\vec{G}_2)
                \widetilde{u}^{\ast}_{-(\vk+\frac{\vk_q}{2}),\ell_3}
                          (\vec{G}_3)
                \widetilde{u}_{-(\vk-\frac{\vk_q}{2})
                           -(\vec{G}_4-\vec{G}_3),\ell_4}(\vec{G}_4)
        \\
                \times
                c^{\dagger}_{\vk+\frac{\vk_q}{2}
                             -(\vec{G}_1-\vec{G}_2) ,\ell_1,\sigma}
                c_{\vk-\frac{\vk_q}{2},\ell_2,\sigma}
                c^{\dagger}_{-(\vk+\frac{\vk_q}{2}),\ell_3,-\sigma}
                c_{-(\vk-\frac{\vk_q}{2})
                   -(\vec{G}_4-\vec{G}_3),\ell_4,-\sigma}.
\end{multline}
By letting
$\vk + \frac{\vk_q}{2} \rightarrow \vk$, and 
$\vk + \frac{\vk_q}{2} \rightarrow \vk'$ we get
\begin{multline} 
\label{V12}
H_I =   \frac{1}{q^2} 
        \sum_{\substack{\vk,\vk',\sigma \\ \ell_1 \cdots \ell_4}}
        \widetilde{V}( \vk - \vk') 
        \sum_{\vec{G}_1 \cdots \vec{G}_4}
                \widetilde{u}^{\ast}_{\vk-(\vec{G}_1-\vec{G}_2),\ell_1}(\vec{G}_1)
                \widetilde{u}_{\vk',\ell_2}(\vec{G}_2)
                \widetilde{u}^{\ast}_{-\vk,\ell_3}(\vec{G}_3)
                \widetilde{u}_{-\vk'-(\vec{G}_4-\vec{G}_3),\ell_4}(\vec{G}_4)
        \\
                \times
                c^{\dagger}_{\vk-(\vec{G}_1-\vec{G}_2) ,\ell_1,\sigma}
                c_{\vk',\ell_2,\sigma}
                c^{\dagger}_{-\vk,\ell_3,-\sigma}
                c_{-\vk'-(\vec{G}_4-\vec{G}_3),\ell_4,-\sigma}.
\end{multline}
To transform this back to the magnetic Brillouin zone we use
\Eqref{u(G)}, in addition to the fact that the combination 
$ e^{-i \vk \cdot \vec{r}_i}
u^{\ast}_{\vk,\ell}(x_i) c^{\dagger}_{\vk,\ell,\sigma} $ has to be
invariant if we let $\vk \rightarrow \vk+\vec{G}$ (as can be seen from
\Eqref{transform}). 
Then
\begin{equation}
        u^{\ast}_{\vk+\vec{G},\ell}(x_i)
        c^{\dagger}_{\vk+\vec{G},\ell,\sigma}
        =
        e^{-i \vec{G} \cdot \vec{r}_i} 
        u^{\ast}_{\vk,\ell}(x_i)
        c^{\dagger}_{\vk,\ell,\sigma}.
\end{equation}
We can then write \Eqref{V12} as
\begin{align}
H_I &=  \frac{1}{2 q^2} 
        \sum_{\substack{\vk,\vk',\sigma \\ \ell_1 \cdots \ell_4}}
        \widetilde{V}( \vk - \vk')
        \begin{aligned}[t]
        \frac{1}{q^2} &
        \sum_{\substack{\vec{G}_1 \cdots \vec{G}_4 \\ \vec{r}_1 \cdots
\vec{r}_4}} 
                u^{\ast}_{\vk,\ell_1}(\vec{r}_1)
                u_{\vk',\ell_2}(\vec{r}_2)
                u^{\ast}_{-\vk,\ell_3}(\vec{r}_3)
                u_{-\vk',\ell_4}(\vec{r}_4)
        \\
                & \times
                e^{-(\vec{G}_1-\vec{G}_2)\cdot\vec{r}_1}
                e^{(\vec{G}_4-\vec{G}_3)\cdot\vec{r}_4}
                e^{ \vec{G}_1 \cdot \vec{r}_1
              -\vec{G}_2 \cdot \vec{r}_2
              +\vec{G}_3 \cdot \vec{r}_3
              -\vec{G}_4 \cdot \vec{r}_4 }
        \\
                & \times
                c^{\dagger}_{\vk,\ell_1,\sigma}
                c_{\vk',\ell_2,\sigma}
                c^{\dagger}_{-\vk,\ell_3,-\sigma}
                c_{-\vk',\ell_4,-\sigma}
        \end{aligned}\\
    &=  \frac{1}{2 q^2}
        \sum_{\substack{\vk,\vk',\sigma \\ \ell_1 \cdots \ell_4}}
        \widetilde{V}( \vk - \vk')
        \begin{aligned}[t]
        \sum_{\substack{\vec{G}_1 \cdots \vec{G}_3 \\ \vec{r}_1 \cdots
\vec{r}_4}} &
                u^{\ast}_{\vk,\ell_1}(\vec{r}_1)
                u_{\vk',\ell_2}(\vec{r}_2)
                u^{\ast}_{-\vk,\ell_3}(\vec{r}_3)
                u_{-\vk',\ell_4}(\vec{r}_4)
        \\
                & \times
                e^{ \vec{G}_2 \cdot (\vec{r}_1-\vec{r}_2) }
                e^{ \vec{G}_3 \cdot (\vec{r}_3-\vec{r}_4) }
                c^{\dagger}_{\vk,\ell_1,\sigma}
                c_{\vk',\ell_2,\sigma}
                c^{\dagger}_{-\vk,\ell_3,-\sigma}
                c_{-\vk',\ell_4,-\sigma}
        \end{aligned} \\ 
     &= \frac{1}{2} 
        \sum_{\substack{\vk,\vk',\sigma \\ \ell_1 \cdots \ell_4}}
        \widetilde{V}( \vk - \vk') 
                \sum_{\vec{r} \vec{r}'} 
                u^{\ast}_{\vk,\ell_1}(\vec{r})
                u_{\vk',\ell_2}(\vec{r})
                u^{\ast}_{-\vk,\ell_3}(\vec{r}')
                u_{-\vk',\ell_4}(\vec{r}')
                c^{\dagger}_{\vk,\ell_1,\sigma}
                c_{\vk',\ell_2,\sigma}
                c^{\dagger}_{-\vk,\ell_3,-\sigma}
                c_{-\vk',\ell_4,-\sigma}
        \nonumber \\
     &= \frac{1}{2}
        \sum_{\substack{\vk,\vk',\sigma \\ \ell_1 \cdots \ell_4}} 
        \underbrace{
                \widetilde{V}( \vk - \vk')  
                \sum_{\vec{G},\vec{G}'} 
                \widetilde{u}^{\ast}_{\vk,\ell_1}(\vec{G})
                \widetilde{u}_{\vk',\ell_2}(\vec{G})
                \widetilde{u}^{\ast}_{-\vk,\ell_3}(\vec{G}')
                \widetilde{u}_{-\vk',\ell_4}(\vec{G}')
        }_{\equiv V^{\ell_1 \ell_3,\ell_2 \ell_4}_{\vk,\vk'}}
                c^{\dagger}_{\vk,\ell_1,\sigma}
                c_{\vk',\ell_2,\sigma}
                c^{\dagger}_{-\vk,\ell_3,-\sigma}
                c_{-\vk',\ell_4,-\sigma}.
\end{align}
{}From \Citeref{Otnes:1997}, we know that the term
$\widetilde{V}( \vk - \vk')$ can be written as
\begin{equation}
        \widetilde{V}( \vk - \vk') 
        = \sum_{\eta=1}^{5}
        \lambda_{\eta} B_{\eta}(\vk) B_{\eta}(\vk'),
\end{equation} 
where $\bvec{\lambda}=(U,V,W,V,W)$, and
\begin{equation}
\begin{split}
        B_1(\vk) &= \sqrt{N} \\
        B_2(\vk) &= \sqrt{N}\left[\cos(\kappa_x)+\cos(\kappa_y)\right]  \\
        B_3(\vk) &= \sqrt{2 N}\cos(\kappa_x) \cos(\kappa_y)  \\
        B_4(\vk) &= \sqrt{N}\left[\cos(\kappa_x)-\cos(\kappa_y)\right] \\
        B_5(\vk) &= \sqrt{2 N}\sin(\kappa_x)\sin(\kappa_y). \\
\end{split}
\end{equation}
The upper limit $\eta\leq 5$ in the sum is due to the
even symmetry of the space wave-function.%
\footnote{%
  The reason for the number $\eta\leq 5$ in the zero-field situation
  is that the gap must be symmetric for $\vec{k} \rightarrow
  -\vec{k}$. A magnetic field destroys time reversal symmetry, and one
  may have to include both even and odd symmetries, then $\eta\leq
  9$.} %
We can then write
\begin{eqnarray}
        V^{\ell_1 \ell_2,\ell'_1 \ell'_2}_{\vk,\vk'}
        =
        \sum_{\vec{G},\vec{G}'} 
        \sum_{\eta=1}^{5}
        \lambda_{\eta} 
        \underbrace{
        \left(\widetilde{u}^{\ast}_{\vk,\ell_1}(\vec{G})
                 \widetilde{u}^{\ast}_{-\vk,\ell_2}(\vec{G}')
                 B_{\eta}(\vk)
        \right)
        }_{\equiv \widetilde{B}^{\ast}_{\eta,G,G'}(\ell_1,\ell_2,\vk)}
        \underbrace{
        \left(\widetilde{u}_{\vk',\ell'_1}(\vec{G})
              \widetilde{u}_{-\vk',\ell'_2}(\vec{G}')
                 B_{\eta}(\vk')
        \right)
        }_{\equiv \widetilde{B}_{\eta,G,G'}(\ell'_1,\ell'_2,\vk')}.
\end{eqnarray}
To simplify notation it is useful to introduce the shorthand notations
$\xi = \{\eta,\vec{G},\vec{G}'\}$, and $\chi = \{\ell_1,\ell_2,\vk\}$
which allows us to write
\begin{equation}
        V_{\chi,\chi'} = \sum_{\xi} \lambda_{\xi} 
        \widetilde{B}^{\ast}_{\xi}(\chi) \widetilde{B}_{\xi}(\chi'),
\end{equation}
where $\lambda_{\xi} \equiv \lambda_{\eta} $.
\end{widetext}



\begin{thebibliography}{28}
\expandafter\ifx\csname natexlab\endcsname\relax\def\natexlab#1{#1}\fi
\expandafter\ifx\csname bibnamefont\endcsname\relax
  \def\bibnamefont#1{#1}\fi
\expandafter\ifx\csname bibfnamefont\endcsname\relax
  \def\bibfnamefont#1{#1}\fi
\expandafter\ifx\csname citenamefont\endcsname\relax
  \def\citenamefont#1{#1}\fi
\expandafter\ifx\csname url\endcsname\relax
  \def\url#1{\texttt{#1}}\fi
\expandafter\ifx\csname urlprefix\endcsname\relax\def\urlprefix{URL }\fi
\providecommand{\bibinfo}[2]{#2}
\providecommand{\eprint}[2][]{\url{#2}}

\bibitem[{\citenamefont{Rasolt and Te{\u{s}}anovi{\'{c}}}(1992)}]{Rasolt:1992}
\bibinfo{author}{\bibfnamefont{M.}~\bibnamefont{Rasolt}} \bibnamefont{and}
  \bibinfo{author}{\bibfnamefont{Z.}~\bibnamefont{Te{\u{s}}anovi{\'{c}}}},
  \bibinfo{journal}{Rev. Mod. Phys.} \textbf{\bibinfo{volume}{64}},
  \bibinfo{pages}{709} (\bibinfo{year}{1992}).

\bibitem[{\citenamefont{Hofstadter}(1976)}]{Hofstadter:1976}
\bibinfo{author}{\bibfnamefont{D.~R.} \bibnamefont{Hofstadter}},
  \bibinfo{journal}{Phys. Rev. B} \textbf{\bibinfo{volume}{14}},
  \bibinfo{pages}{2239} (\bibinfo{year}{1976}).

\bibitem[{\citenamefont{Otnes and Sudb{\o}}(1999)}]{Otnes:1997}
\bibinfo{author}{\bibfnamefont{E.}~\bibnamefont{Otnes}} \bibnamefont{and}
  \bibinfo{author}{\bibfnamefont{A.}~\bibnamefont{Sudb{\o}}},
  \bibinfo{journal}{Int. journal of mod. phys. B}
  \textbf{\bibinfo{volume}{13}}, \bibinfo{pages}{1579} (\bibinfo{year}{1999}),
  \eprint{cond-mat/9707225}.

\bibitem[{\citenamefont{Harper}(1955)}]{Harper:1955}
\bibinfo{author}{\bibfnamefont{P.~G.} \bibnamefont{Harper}},
  \bibinfo{journal}{Proc. Phys. Soc. London A} \textbf{\bibinfo{volume}{68}},
  \bibinfo{pages}{874} (\bibinfo{year}{1955}).

\bibitem[{\citenamefont{Brown}(1964)}]{Brown:1964}
\bibinfo{author}{\bibfnamefont{E.}~\bibnamefont{Brown}},
  \bibinfo{journal}{Phys. Rev.} \textbf{\bibinfo{volume}{133}},
  \bibinfo{pages}{1038} (\bibinfo{year}{1964}).

\bibitem[{\citenamefont{Brown}(1968)}]{Brown:1968b}
\bibinfo{author}{\bibfnamefont{E.}~\bibnamefont{Brown}}, in
  \emph{\bibinfo{booktitle}{Solid State Physics}}, edited by
  \bibinfo{editor}{\bibfnamefont{H.}~\bibnamefont{Erenreich}},
  \bibinfo{editor}{\bibfnamefont{F.}~\bibnamefont{Seitz}}, \bibnamefont{and}
  \bibinfo{editor}{\bibfnamefont{D.}~\bibnamefont{Turnbull}}
  (\bibinfo{publisher}{Academic, New York}, \bibinfo{year}{1968}),
  vol.~\bibinfo{volume}{22}, pp. \bibinfo{pages}{313--408}.

\bibitem[{\citenamefont{Zak}(1964{\natexlab{a}})}]{Zak:1964a}
\bibinfo{author}{\bibfnamefont{J.}~\bibnamefont{Zak}}, \bibinfo{journal}{Phys.
  Rev.} \textbf{\bibinfo{volume}{134}}, \bibinfo{pages}{1602}
  (\bibinfo{year}{1964}{\natexlab{a}}).

\bibitem[{\citenamefont{Zak}(1964{\natexlab{b}})}]{Zak:1964b}
\bibinfo{author}{\bibfnamefont{J.}~\bibnamefont{Zak}}, \bibinfo{journal}{Phys.
  Rev.} \textbf{\bibinfo{volume}{134}}, \bibinfo{pages}{1607}
  (\bibinfo{year}{1964}{\natexlab{b}}).

\bibitem[{\citenamefont{Zak}(1964{\natexlab{c}})}]{Zak:1964d}
\bibinfo{author}{\bibfnamefont{J.}~\bibnamefont{Zak}}, \bibinfo{journal}{Phys.
  Rev.} \textbf{\bibinfo{volume}{136}}, \bibinfo{pages}{1647}
  (\bibinfo{year}{1964}{\natexlab{c}}).

\bibitem[{\citenamefont{Zak}(1965)}]{Zak:1965}
\bibinfo{author}{\bibfnamefont{J.}~\bibnamefont{Zak}}, \bibinfo{journal}{Phys.
  Rev.} \textbf{\bibinfo{volume}{139}}, \bibinfo{pages}{1159}
  (\bibinfo{year}{1965}).

\bibitem[{\citenamefont{Dana and Zak}(1983)}]{Dana:1983}
\bibinfo{author}{\bibfnamefont{I.}~\bibnamefont{Dana}} \bibnamefont{and}
  \bibinfo{author}{\bibfnamefont{J.}~\bibnamefont{Zak}},
  \bibinfo{journal}{Phys. Rev. B} \textbf{\bibinfo{volume}{28}},
  \bibinfo{pages}{811} (\bibinfo{year}{1983}).

\bibitem[{\citenamefont{Dana and Zak}(1985)}]{Dana:1985}
\bibinfo{author}{\bibfnamefont{I.}~\bibnamefont{Dana}} \bibnamefont{and}
  \bibinfo{author}{\bibfnamefont{J.}~\bibnamefont{Zak}},
  \bibinfo{journal}{Phys. Rev. B} \textbf{\bibinfo{volume}{32}},
  \bibinfo{pages}{3612} (\bibinfo{year}{1985}).

\bibitem[{\citenamefont{Obermair and Schellnhuber}(1981)}]{Obermair:1981a}
\bibinfo{author}{\bibfnamefont{G.~M.} \bibnamefont{Obermair}} \bibnamefont{and}
  \bibinfo{author}{\bibfnamefont{H.-J.} \bibnamefont{Schellnhuber}},
  \bibinfo{journal}{Phys. Rev. B} \textbf{\bibinfo{volume}{23}},
  \bibinfo{pages}{5185} (\bibinfo{year}{1981}).

\bibitem[{\citenamefont{Obermair et~al.}(1981)\citenamefont{Obermair,
  Schellnhuber, and Rauh}}]{Obermair:1981b}
\bibinfo{author}{\bibfnamefont{G.~M.} \bibnamefont{Obermair}},
  \bibinfo{author}{\bibfnamefont{H.-J.} \bibnamefont{Schellnhuber}},
  \bibnamefont{and} \bibinfo{author}{\bibfnamefont{A.}~\bibnamefont{Rauh}},
  \bibinfo{journal}{Phys. Rev. B} \textbf{\bibinfo{volume}{23}},
  \bibinfo{pages}{5191} (\bibinfo{year}{1981}).

\bibitem[{\citenamefont{Thouless et~al.}(1982)\citenamefont{Thouless, Kohmoto,
  Nightingale, and den Nijs}}]{Thouless:1982}
\bibinfo{author}{\bibfnamefont{D.~J.} \bibnamefont{Thouless}},
  \bibinfo{author}{\bibfnamefont{M.}~\bibnamefont{Kohmoto}},
  \bibinfo{author}{\bibfnamefont{M.~P.} \bibnamefont{Nightingale}},
  \bibnamefont{and} \bibinfo{author}{\bibfnamefont{M.}~\bibnamefont{den Nijs}},
  \bibinfo{journal}{Phys. Rev. Lett.} \textbf{\bibinfo{volume}{49}},
  \bibinfo{pages}{405} (\bibinfo{year}{1982}).

\bibitem[{\citenamefont{Schellnhuber}(1982)}]{Schellnhuber:1982}
\bibinfo{author}{\bibfnamefont{H.-J.} \bibnamefont{Schellnhuber}},
  \bibinfo{journal}{Phys. Rev. B} \textbf{\bibinfo{volume}{25}},
  \bibinfo{pages}{2358} (\bibinfo{year}{1982}).

\bibitem[{\citenamefont{Hasegawa et~al.}(1989)\citenamefont{Hasegawa, Lederer,
  Rice, and Wiegmann}}]{Hasegawa:1989}
\bibinfo{author}{\bibfnamefont{Y.}~\bibnamefont{Hasegawa}},
  \bibinfo{author}{\bibfnamefont{P.}~\bibnamefont{Lederer}},
  \bibinfo{author}{\bibfnamefont{T.~M.} \bibnamefont{Rice}}, \bibnamefont{and}
  \bibinfo{author}{\bibfnamefont{P.~B.} \bibnamefont{Wiegmann}},
  \bibinfo{journal}{Phys. Rev. Lett.} \textbf{\bibinfo{volume}{63}},
  \bibinfo{pages}{907} (\bibinfo{year}{1989}).

\bibitem[{\citenamefont{Barelli et~al.}(1996)\citenamefont{Barelli, Bellissard,
  Jacquod, and Shepelyansky}}]{Barelli:1996}
\bibinfo{author}{\bibfnamefont{A.}~\bibnamefont{Barelli}},
  \bibinfo{author}{\bibfnamefont{J.}~\bibnamefont{Bellissard}},
  \bibinfo{author}{\bibfnamefont{P.}~\bibnamefont{Jacquod}}, \bibnamefont{and}
  \bibinfo{author}{\bibfnamefont{D.~L.} \bibnamefont{Shepelyansky}},
  \bibinfo{journal}{Phys. Rev. Lett.} \textbf{\bibinfo{volume}{77}},
  \bibinfo{pages}{4752} (\bibinfo{year}{1996}).

\bibitem[{\citenamefont{Shepelyansky}(1996)}]{Shepelyansky:1996}
\bibinfo{author}{\bibfnamefont{D.~L.} \bibnamefont{Shepelyansky}},
  \bibinfo{journal}{Phys. Rev. B} \textbf{\bibinfo{volume}{54}},
  \bibinfo{pages}{14896} (\bibinfo{year}{1996}), \eprint{cond-mat/9609134}.

\bibitem[{\citenamefont{Ahn and Chang}(1997)}]{Anh:1997}
\bibinfo{author}{\bibfnamefont{K.-H.} \bibnamefont{Ahn}} \bibnamefont{and}
  \bibinfo{author}{\bibfnamefont{K.~J.} \bibnamefont{Chang}},
  \bibinfo{journal}{Phys. Rev. B} \textbf{\bibinfo{volume}{56}},
  \bibinfo{pages}{12772} (\bibinfo{year}{1997}).

\bibitem[{\citenamefont{Doh and Salk}(1998)}]{Doh:1998}
\bibinfo{author}{\bibfnamefont{H.}~\bibnamefont{Doh}} \bibnamefont{and}
  \bibinfo{author}{\bibfnamefont{S.-H.~S.} \bibnamefont{Salk}},
  \bibinfo{journal}{Phys. Rev. B} \textbf{\bibinfo{volume}{57}},
  \bibinfo{pages}{1312} (\bibinfo{year}{1998}).

\bibitem[{\citenamefont{Hong and Salk}(1999)}]{Hong:1999}
\bibinfo{author}{\bibfnamefont{S.-P.} \bibnamefont{Hong}} \bibnamefont{and}
  \bibinfo{author}{\bibfnamefont{S.-H.~S.} \bibnamefont{Salk}},
  \bibinfo{journal}{Phys. Rev. B} \textbf{\bibinfo{volume}{60}},
  \bibinfo{pages}{9550} (\bibinfo{year}{1999}).

\bibitem[{\citenamefont{Hong et~al.}(2000)\citenamefont{Hong, Lee, and
  Salk}}]{Hong:2000}
\bibinfo{author}{\bibfnamefont{S.-P.} \bibnamefont{Hong}},
  \bibinfo{author}{\bibfnamefont{S.-S.} \bibnamefont{Lee}}, \bibnamefont{and}
  \bibinfo{author}{\bibfnamefont{S.-H.~S.} \bibnamefont{Salk}},
  \bibinfo{journal}{Phys. Rev. B} \textbf{\bibinfo{volume}{62}},
  \bibinfo{pages}{14880} (\bibinfo{year}{2000}).

\bibitem[{\citenamefont{Mierzejewski and Maska}(1999)}]{Mierzejewski:1999prb}
\bibinfo{author}{\bibfnamefont{M.}~\bibnamefont{Mierzejewski}}
  \bibnamefont{and} \bibinfo{author}{\bibfnamefont{M.}~\bibnamefont{Maska}},
  \bibinfo{journal}{Phys. Rev. B} \textbf{\bibinfo{volume}{60}},
  \bibinfo{pages}{6300} (\bibinfo{year}{1999}).

\bibitem[{\citenamefont{Maska and Mierzejewski}(2000)}]{Maska:2000}
\bibinfo{author}{\bibfnamefont{M.~M.} \bibnamefont{Maska}} \bibnamefont{and}
  \bibinfo{author}{\bibfnamefont{M.}~\bibnamefont{Mierzejewski}}
  (\bibinfo{year}{2000}), \eprint{cond-mat/0005142}.

\bibitem[{\citenamefont{Miyazaki et~al.}(1999)\citenamefont{Miyazaki, Kishigi,
  and Hasegawa}}]{Miyazaki:1999cm}
\bibinfo{author}{\bibfnamefont{M.}~\bibnamefont{Miyazaki}},
  \bibinfo{author}{\bibfnamefont{K.}~\bibnamefont{Kishigi}}, \bibnamefont{and}
  \bibinfo{author}{\bibfnamefont{Y.}~\bibnamefont{Hasegawa}}
  (\bibinfo{year}{1999}), \eprint{cond-mat/9901194}.

\bibitem[{\citenamefont{Han. et~al.}(1994)\citenamefont{Han., Thouless,
  Hiramoto, and Kohmoto}}]{Han:1994}
\bibinfo{author}{\bibfnamefont{J.~H.} \bibnamefont{Han.}},
  \bibinfo{author}{\bibfnamefont{D.~J.} \bibnamefont{Thouless}},
  \bibinfo{author}{\bibfnamefont{H.}~\bibnamefont{Hiramoto}}, \bibnamefont{and}
  \bibinfo{author}{\bibfnamefont{M.}~\bibnamefont{Kohmoto}},
  \bibinfo{journal}{Phys. Rev. B} \textbf{\bibinfo{volume}{50}},
  \bibinfo{pages}{11365} (\bibinfo{year}{1994}).

\bibitem[{\citenamefont{Peierls}(1933)}]{Peierls:1933}
\bibinfo{author}{\bibfnamefont{R.~E.} \bibnamefont{Peierls}},
  \bibinfo{journal}{Z. Physik} \textbf{\bibinfo{volume}{80}},
  \bibinfo{pages}{763} (\bibinfo{year}{1933}).

\end{thebibliography}
\end{document}